\documentclass[]{JHEP3}
\usepackage{graphics}
\usepackage{epsfig}

\def\homega{\omo}
\def\nc{n}

\def\bg{\bar{g}}
\def\bnabla{\bar{\nabla}}

\def\lb{\left(}
\def\rb{\right)}
\def\ls{\left[}
\def\rs{\right]}
\def\lc{\left\{}
\def\td{\tilde}
\def\rc{\right\}}
\def\tphi{\tilde\phi}

\newcommand{\be}{\begin{equation}}
\newcommand{\ee}{\end{equation}}
\newcommand{\bea}{\begin{eqnarray}}
\newcommand{\eea}{\end{eqnarray}}
\newcommand{\lsim}{\mbox{\raisebox{-.6ex}{~$\stackrel{<}{\sim}$~}}}

\newcommand{\mx}{\mbox}
\newcommand{\mt}{\mathtt}

\newcommand{\p}{\partial}

\newcommand{\al}{\alpha}

\newcommand{\te}{\theta}

\newcommand{\om}{\omega}

\newcommand{\ti}{\widetilde}

\newcommand{\ms}{{m}}
\newcommand{\ns}{{n}}
\newcommand{\Ds}{n}
\newcommand{\omo}{\hat{\om}}
\newcommand{\2}{\frac{1}{2}}
\newcommand{\3}{\textstyle{1\over 3}}

\newcommand{\ra}{\rightarrow}

\newcommand{\Ra}{\Rightarrow}

\newcommand{\sss}{\scriptscriptstyle}

\title{Moduli Stabilization in Brane Gas Cosmology with Superpotentials }
\author{Aaron Berndsen, Tirthabir Biswas, and James M.\ Cline\\
Physics Department, McGill University, Montr\'eal, Qu\'ebec, Canada H3A
2T8\\
E-mail: \email{aberndsen@physics.mcgill.ca}, 
\email{tirtho@physics.mcgill.ca}, \email{jcline@physics.mcgill.ca}}

\preprint{}
\abstract{
In the context of brane gas cosmology in  superstring theory, 
we show why it is impossible to simultaneously stabilize   the
dilaton and the radion with a general gas of strings (including massless modes) and D-branes.
Although this requires invoking a different mechanism to stabilize
these moduli fields, we find that the brane gas can still play a
crucial role in the early universe in assisting moduli stabilization.
We show that a modest energy density of
specific types of brane gas can solve the overshoot problem that
typically afflicts potentials arising from gaugino condensation.}
\begin{document}
\section{Introduction}
\label{sec:intro}
A major success of brane gas cosmology (BGC) is the utilization of stringy
effects to explain the origin of the hierarchy of dimensions. In the
seminal proposal of Brandenberger and Vafa~\cite{branvafa}, it was argued that
in the early universe all directions could fluctuate about the self-dual
radius due to the presence of both winding and momentum modes. The
argument asserts that strings will generically intersect in
(3+1)-dimensional subspaces, so that such a subspace will lose its winding and
subsequently expand into the large directions we observe today.
 This scenario was mathematically realized by Tseytlin and Vafa in the
 context of dilaton gravity~\cite{tseyvaf}, and has since been extended to
include the effects of a gas of $Dp$-branes, where Alexander, Easson,
and Brandenberger~\cite{abe} argued that such a gas would result in a
hierarchy of extra dimensions. Namely the original $9$-dimensional
spatial manifold should decompactify into a hierarchical product space of
$\mathcal{T}_4\times\mathcal{T}_2\times\mathcal{T}_3$. 

Subsequent investigations indicate that wound strings provide a
mechanism for isotropization \cite{branwat_isot} and stabilization
\cite{branwat_stab} of the the compact dimensions, and that the
mechanism works on toroidal orbifolds~\cite{egj}. The framework for
these results is usually the low energy effective action of type IIA
string theory, where the salient differences from general relativity
are a massless dilaton and the dynamics of extra dimensions. These
differences lead to the result that negative pressure in the compact
directions, due to wound strings,  results in contraction,  not
acceleration.  The dilaton is assumed to have no potential other than
that which is induced by its coupling to the bulk string frame
Lagrangian $e^{-2\phi}(R + (\nabla\phi)^2)$ and possible D-brane
sources; most successes of BGC rely on the dynamical running of the
dilaton toward weak string coupling, $g_s\ll 1$. 

On the other hand one would like to stabilize the dilaton at a value
where $g_s$ is still large enough to be consistent with  gauge
coupling unification \cite{GUT}. Moreover if $g_s$ becomes too small, the
interactions between strings become too weak to allow the
annihilation of winding modes in three dimensions, where the space
should be allowed to grow  \cite{egjk1}. (See also \cite{anupam1} for a discussion similar issues.) Rather there is only a
window of finely-tuned initial conditions consistent with three
dimensions ultimately growing  to be large.   A third reason that the
dilaton must not continue to roll to arbitrarily small values is the
constraint from fifth force experiments and null searches
for time variation of physical constants \cite{5th}; these
preclude the dilaton from continuing to evolve at late times. 

For these reasons it is imperative to reconcile brane gas cosmology
with the stabilization of  both the radion and the dilaton. In
\cite{batwat} it was shown that using just the string winding and
momentum modes this is not possible. We therefore first investigate 
whether by including more general string and brane states one can 
achieve such a stabilization. (A similar but less general analysis
has been done in \cite{arakay}.)

Our result, in the context of superstring theories, is that a
general gas of D-branes and strings cannot stabilize both moduli, although they can
stabilize one linear combination of them.  However, the brane gas can
still play an important role in the process of stabilizing  both
fields, due to the overshoot problem \cite{brustein}. A much-studied
mechanism for stabilizing the radion involves adding racetrack
potentials coming  from gaugino condensation (and possibly an
antibrane \cite{kklt}). The Minkowski minimum of these potentials is
typically separated from a runaway (decompactification) direction by
a very small barrier, which would always be overcome by the inertia
of the fields if their initial conditions were not finely tuned to be
close to the desired minimum.  One of our main observations is that a
gas of brane winding modes  can very robustly solve this problem by
slowing down the modulus as it rolls down its steep potential.

\par Our plan is as follows: in Section~\ref{strtoein} we describe
the BGC scenario and motivate the dimensional reduction procedure to
obtain a $d$-dimensional theory of gravity with two scalar fields
(the dilaton and radion), with an effective potential coming from the
brane gas. In Section \ref{sec:res} we discuss some of the  features
of the effective potential; namely, we show that provided the dilaton is stabilized by some other mechanism, branes can stabilize all 
the extra dimensions. Section \ref{sec3} presents our no-go  theorem
showing that under the given assumptions, there exists an
unstabilized direction in the moduli space of the dilaton and radion
no matter what modes are included in the gas of D-branes and strings. In particular we also show that the presence of massless F-string modes do not help in lifting the runaway direction. In section
\ref{sec4} we
consider the combined effect of the brane gas with a superpotential,
such as would arise from gaugino condensation and antibranes, and
show that the brane gas can provide a remedy for the overshoot
problem.   We
give our conclusions in section \ref{sec6}.  Technical details are given in
the appendices.

\section{Effective Brane Gas Cosmology}
\label{sec:BGC}
\subsection{Supergravity coupled to Strings and Brane Sources}
\label{strtoein}
A starting point for BGC is type~IIA
string theory compactified on a 9-dimensional toroidal background, which may be
thought of as the result of compactifying $M$-theory on $S^1$. The
low-energy bulk effective action of this theory is given by
\be
\label{eq:ttap}
S_{IIa}=\frac{1}{2\kappa^2}\int d^Dx\sqrt{-G}\,e^{-2\phi}\lb
R+4G^{MN}\nabla_M\phi\nabla_N\phi-\frac{1}{12}H_{\mu\nu\alpha}H^{\mu\nu\alpha}\rb\ ,
\ee
where $G$ is the determinant of the ten-dimensional background metric
$G_{\mu\nu}$, $\phi$ is the dilaton, $H$ is the field strength
corresponding to the bulk antisymmetric tensor field $B_{\mu\nu}$, and
$\kappa$ is the D-dimensional Newton's constant. For simplicity we
ignore any flux contributions, and take $H=0$. We envision
this analysis to apply in the late-time era of BGC 
\cite{branvafa,abe,egj,bek},
an epoch
where the extra, compact dimensions are expected to be
isotropized~\cite{branwat_isot}, and winding modes in the large
directions have annihilated. Thus, we consider a spacetime
consisting of a flat, $d$-dimensional FRW universe, and an isotropic compact
subspace of $\nc$ extra dimensions
\bea
ds^2&=&G_{MN}dX^MdX^N=g_{\mu\nu}dx^\mu dx^\nu+b^2(t)\,\gamma_{mn}dy^m dy^n\\
&=&-dt^2+a^2(t)\,dx_idx^i+b^2(t)\,dy_mdy^m,\
i\in\{1,\ldots,d\},m\in\{1,\ldots,\nc\}\ 
\label{eq:metric}
\eea
where $y^m$ are the coordinates of the $\nc$ extra dimensions. The total action
comprises the above bulk action~(\ref{eq:ttap}) and the action of all matter
present. Sources are included by adding  matter terms for
both the strings ($\rho_s$) and Dp-branes ($\rho_p$). Owing to the different
world-sheet couplings between the dilaton and the branes and strings, the
matter action has the form
\bea
S_m&=&-\int d^Dx\sqrt{-G}\lb\rho_s+ e^{-\phi}\rho_p\rb\label{eq:mat}\\
T_{MN}&=&-\frac{2}{\sqrt{-G}}\frac{\delta S_m}{\delta G^{MN}}\ .
\eea
\par
We continue the construction of late-time BGC by considering separate
species of strings and branes, each possibly having excited momentum
(in the case of branes also known as  ``vibrational
modes''~\cite{kaya_vol}) 
in the large or compact subspaces, but
having winding modes only along the compact directions. Then one can
show (see appendix \ref{app:exps}) that the stress energy tensor for
the strings and branes simplifies to 
\bea
- T^0_0=\rho_s+e^{-\phi}\rho_p&=&\sum_i\ls \rho_ie^{-\alpha_i\phi}a^{-d(1+\omega_i)}b^{-\nc(1+\homega_i)}\rs\label{eq:t00s}\\
T^a_b=P\,\delta^a_b&=&\sum_i\omega_i\ls\rho_ie^{-\alpha_i\phi}a^{-d(1+\omega_i)}b^{-\nc(1+\homega_i)}\rs\delta^a_b\\
 T^m_n=p\,\delta^m_n&=&\sum_i\homega_i\ls\rho_ie^{-\alpha_i\phi}a^{-d(1+\omega_i)}b^{-\nc(1+\homega_i)}\rs\delta^m_n\label{eq:tmms}.
\eea
In the preceding expressions the summation is performed over the
relevant modes contributing to the gas of strings and branes, $P$ and
$p$ being the sum-total pressure along the large and compact
directions respectively. $\alpha_i=0$ for string sources,
$\alpha_i=1$ for brane sources,  and  $\rho_i$ is the initial energy
density for a particular mode, with effective equation of state
$p_i=\homega_i\rho_i$, $P_i=\omega_i\rho_i$. The values of $\om$ and
$\homega$ depend on the specific type of mode, dimensionality of the
branes and the  number of large and extra dimensions  (see table
\ref{tab:exps}), but the important thing is that all the known modes
can be described by these quantities.
\par
Variation of the action~(\ref{eq:ttap}) together with the matter action~(\ref{eq:mat})
and metric ansatz~(\ref{eq:metric}) results in the system of equations
\bea
-d\lb\frac{\dot a(t)}{a(t)}\rb^2-\nc\lb\frac{\dot b(t)}{b(t)}\rb^2+\dot\varphi^2=&e^\varphi E&\hspace{1cm}\label{eq:bgc1}\\
\frac{d}{dt}\lb\frac{\dot a(t)}{a(t)}\rb-\dot\varphi\frac{\dot a(t)}{a(t)}=&\frac{1}{2}e^\varphi P&\hspace{1cm}\label{eq:bgc2}\\
\frac{d}{dt}\lb\frac{\dot b(t)}{b(t)}\rb-\dot\varphi\frac{\dot b(t)}{b(t)}=&\frac{1}{2}e^\varphi p&\hspace{1cm}\label{eq:bgc3}\\
\ddot\varphi-d\lb\frac{\dot a(t)}{a(t)}\rb^2-\nc\lb\frac{\dot
  b(t)}{b(t)}\rb^2=&\frac{1}{2}e^\varphi E\ ,&\hspace{1cm}\label{eq:bgc4}
\eea
where
we have introduced the shifted dilaton as $\varphi\equiv
2\phi-d\frac{\dot a(t)}{a(t)}-n\frac{\dot b(t)}{b(t)}$ (recall that $d=3$ and
$n=6$), and a dot
denotes differentiation with respect to time. Eqs.\ (\ref{eq:bgc1}-\ref{eq:bgc4})
are the  string frame, or dilaton-gravity, equations of motion. 
Equation~(\ref{eq:bgc1}) is
the $0$-$0$ Einstein equation; notice that in the string frame the kinetic
term for the (shifted) dilaton contributes to the energy with apparently the
wrong sign---this is due to the nonminimal coupling between the Ricci
Scalar and the dilaton. The spatial components of the Einstein
equations~(\ref{eq:bgc2}-\ref{eq:bgc3}) show that the acceleration of
the scale factor is
proportional to the pressure, and thus the negative-pressure winding modes
lead to contraction---this is the key ingredient of the
Brandenberger-Vafa mechanism. Eq.\ (\ref{eq:bgc4}) is the dilaton equation of
motion.  Equation~(\ref{eq:bgc1}) is not dynamical, but is rather an equation 
of constraint, which  can be used to determine the initial dilaton velocity
to be
\be
\dot\varphi=\pm\sqrt{e^{\varphi}E+d\lb\frac{\dot
    a(t)}{a(t)}\rb^2+\nc\lb\frac{\dot b(t)}{b(t)}\rb^2}\ .
\ee
It is customary to choose the negative
solution since the string coupling $g_s = e^\phi$ then evolves toward
weak coupling, where a perturbative description is valid.
Since all the terms under the square root are positive, the
dilaton cannot bounce.
\par
The rolling of the dilaton, although important for the BV mechanism and the 
stabilization of the moduli fields, may also be deleterious to the BGC
scenario. In~\cite{egjk1}, Easther, Greene, Jackson, and Kabat show that if the
dilaton rolls too quickly, winding-mode annihilation may be suppressed, so that
dynamical evolution leading to three large spatial dimensions is not favoured.
The rolling of the dilaton also implies evolution of
volume of the compact space.
A conformal transformation on the metric may absorb the
$\phi$-$R$ coupling term, but this means the Einstein frame scale factors get 
additional time dependence from $\phi(t)$. This problem has typically been
set aside (on the assumption that the dilaton will be stabilized
at a later time) in discussions of 
stabilization of the extra dimensions~\cite{branwat_stab}. However, in order to have a
complete and consistent picture in the framework of brane gas cosmology one indeed
needs to address the issue of stabilizing the dilaton along with radion stabilization,
and this is what we devote the next few subsections to. 
\par
The preceding observations also stress the utility of viewing gravity 
from the point of view of the four-dimensional Einstein frame, 
which is more intuitive than the 10D string frame. 
An effective four-dimensional
action is achieved by conformally
absorbing the dilaton, integrating out the extra dimensions, and
performing a second conformal transformation to absorb the scale
factor of the extra dimensions. The result is a minimally-coupled
theory of BGC, where the original string and brane sources act as an effective
potential for both the radion and dilaton fields. This approach was first advocated in \cite{zhuk} to study stabilization of extra dimensions in the presence of hydrodynamical fluids and  was used to study string winding and momentum modes in \cite{batwat}. We
now generalize the analysis to include all possible string and brane sources.

\subsection{Effective Potential}
\label{effact}
Upon performing dimensional reduction on both the string
and brane sources, we obtain general relativity coupled
to two scalar fields, the dilaton and radion, with an
effective potential coming from the brane gas. As outlined in
Appendix~\ref{app:dimred}, a string/brane source whose energy density behaves as
$\rho=\rho_i e^{-\alpha_i\phi} a^{-d(1+\omega_i)}b^{-n(1+\homega_i)}$, with equations
of state $\omega_i$ and $\homega_i$ in the $d$ large and $\nc$ compact directions
respectively, provides an effective potential in $d+1$ dimensions 
\bea
\label{eq:veffst}
V_{{\rm eff}, i}&=&
\rho_i\,e^{2\nu_i\psi}\,e^{2\mu_i\varphi}\,\bar{a}^{-d(1+\omega_i)} \nonumber\\
\nu_i &=& \frac12{\lb-\homega_i+\frac{d}{d-1}\lb\omega_i-\frac1d\rb\rb
\sqrt{\frac{(d-1)\nc}{(d+\nc-1)}}}\,\nonumber\\
\mu_i &=& \frac12{\lb-d\omega_i-\nc\homega_i+1-\alpha_i\frac{d+\nc-1}{2}\rb
\sqrt{\frac{1}{d+\nc-1}}}\,
\eea
This is expressed in terms of the canonically normalized moduli
$\psi$ and dilaton $\varphi$ fields, and  the Einstein-frame scale
factor $\bar a$ of the $d$ large directions. $\alpha_i$ parametrizes
string ($\alpha_i=0$) or brane ($\alpha_i=1$) contributions, and
$\rho_i$ is the initial energy density of the $i$th component of the
brane gas. We work in Planck units, $M_{pl}^{-2}=8\pi G_N=1$. The net
effective potential will comprise several contributions of the
form~(\ref{eq:veffst}), depending on the type of excited modes;
Appendix~\ref{app:exps}  discusses the equations of state, and the
coefficients $\mu_i,\ \nu_i$ for the various string and brane
sources and the results are  summarized in Table 1.

\TABLE{\centering
\begin{tabular}{|cccccc|}
\hline
source&$E\propto a^{-d\omega}b^{-n\homega}$&$\omega$&$\homega$&$\mu_i$&$\nu_i$\\\hline\hline
\multicolumn{6}{|c|}{$d=3$}\\
\hline
general
string&$a^{-d\omega}b^{-n\homega}$&$\omega$&$\homega$&${-3\omega-\nc\homega+1\over\sqrt{{4(\nc+2)}}}$&
$-{\left(\homega+\frac12\left(1-3\omega\right)\right)\sqrt{\nc}\over\sqrt{{2(\nc+2)}}}$\\
general
brane&$a^{-d\omega}b^{-n\homega}$&$\omega$&$\homega$&$-{3\omega+n\homega+\frac{n}2\over\sqrt{{4(\nc+2)}}}$
&$-{\left(\homega+\frac12\left(1-3\omega\right)\right)\sqrt{\nc}\over\sqrt{{2(\nc+2)}}}$\\
wound
string&$a^0b^1$&0&$\frac{-1}{\nc}$&$\frac{1}{\sqrt{\nc+2}}$&${1-\frac\nc2\over\sqrt{{2\nc(\nc+2)}}}$\\
wound
brane&$a^0b^p$&0&$\frac{-p}{\nc}$&$\frac{(p-\frac\nc2)}{2\sqrt{\nc+2}}$&${p-\frac\nc2\over\sqrt{{2\nc(\nc+2)}}}$\\
string
momentum&$a^0b^{-1}$&0&$\frac{1}{\nc}$&0&$-\sqrt{\frac{\nc+2}{8\nc}}$\\
brane
momentum&$a^0b^{-1}$&0&$\frac{1}{\nc}$&$-\frac14\sqrt{\nc+2}$&$-\sqrt{\frac{\nc+2}{8\nc}}$\\
\hline\hline
\multicolumn{6}{|c|}{$d=3,\ \nc=6$}\\\hline
wound
string&$a^0b^1$&0&$\frac{-1}{6}$&$\frac{1}{\sqrt{8}}$&$-\frac{1}{2\sqrt{6}}$\\
wound
brane&$a^0b^p$&0&$\frac{-p}{\nc}$&$\frac{p-3}{2\sqrt{8}}$&$\frac{p-3}{4\sqrt{6}}$\\
string
momentum&$a^0b^{-1}$&0&$\frac{1}{\nc}$&0&$\frac{-1}{\sqrt{6}}$\\
brane
momentum&$a^0b^{-1}$&0&$\frac{1}{\nc}$&$-\frac{1}{\sqrt{2}}$&$\frac{-1}{\sqrt{6}}$\\\hline
\end{tabular} \label{tab:exps}\caption{A summary of the radion and
dilaton couplings in the effective potential due to various species of gas. The
spatial background consists of $d$ large and $n$ compact directions,
with equations of states $\omega$ and $\homega$ respectively. }
}

\subsection{Radion Stabilization}
\label{sec:res}To understand the effects of string and brane sources in
late-time BGC, we now specialize to the case of three large directions ($d=3$)
with winding modes only in the compact dimensions. First suppose  that brane sources
are not present, so the effective potential for the system is given by
contributions from strings alone---this emulates the setup
of \cite{branvafa,branwat_stab,batwat,bek,bercli}. Three representative
species of strings are
considered, namely, $W$: strings with winding numbers
in the compact direction
($\omega=0$, $\homega=-\frac1n$), $M_6$: momentum excitations in the compact directions
($\omega=0$, $\homega=\frac1n$), and  $M_3$: momentum in the large
directions ($\omega=\frac1d$, $\homega=0$). Summing contributions~(\ref{eq:veffst}),
we obtain
\bea
\label{eq:veffs3}
V_s(\bar a,\varphi,\psi)&=& \rho_{\sss W}e^{(1-
\frac{\nc}{2})\sqrt{B}\psi}e^{\sqrt{\frac
A2}\varphi}\bar{a}^{-3}+\rho_{\sss M_6}e^{-(1+\frac{\nc}{2})\sqrt{B}\psi}
\bar{a}^{-3}+\rho_{\sss M_3}\bar{a}^{-4}\ ,
\eea
where $\rho_{\sss W}$, $\rho_{\sss M_3}$, $\rho_{\sss M_6}$ parametrize  the initial
energy densities of the three kinds of components,  $B=\frac{2}{\nc(\nc+2)}$, and
$A=\frac{2}{\nc+2}$. Let us assume that the dilaton has been stabilized by an
external potential and consider the effect of the string gases on the unstabilized
radion.  Taking the dilaton VEV to be $\phi=0$ and ignoring the
$M_3$ momentum modes, which anyway gets redshifted by the expansion of the universe,
the string gas effective potential (\ref{eq:veffs3}) becomes
\bea
V_s(\bar a,\psi)&=&\bar{a}^{-3}\left[\rho_Ne^{(1-
\frac{\nc}{2})\sqrt{B}\psi}+\rho_Me^{-(1+\frac{\nc}{2})\sqrt{B}\psi}\right]\ .
\eea
Battefeld and Watson point out~\cite{batwat} that this is a stable potential for
$\psi$  only if the number of extra dimensions is $\nc=1$, in which case it reduces to
$V(\bar a, \psi=0)\sim\frac{1}{\bar a^3}$. This can be considered a source of dark
matter, similar to the string-inspired example of Gubser and Peebles \cite{gubpee}.
However, in the case of $n>2$,~\cite{batwat} points out that  the effective potential behaves
as $V(\bar a,\psi)\sim {e^{-a\psi}}/{\bar a^3}$, so that the
radion also runs  away to $\infty$. Since $\nc=6$, one sees   that the presence of
strings cannot stabilize the dilaton or the radion. 

We note that in \cite{Watson,subodh} massless string  states were invoked to
obtain stabilization of the moduli. However, the former are not present in
the type II string (being removed by the GSO projection). Although, they are
 present at the self-dual radius in the heterotic
string,\footnote{We thank Subodh Patil for discussions on this
point.}  additionally, \cite{subodh}  requires quantized modes of the
D-string to achieve complete stabilization of all moduli.  It is not
clear to us that the D-string can be quantized in the same way as the
fundamental string.

\par

Let us therefore consider whether extending the analysis of \cite{batwat} to the
case of general brane sources can solve the problem of moduli stabilization. Consider
the contributions to $V_{\rm eff}$ coming from  $p$-branes wrapping the compact
dimensions ($\omega=0$, $\homega=-\frac pn$, denoted $\td N$), and  momentum modes in
the compact dimensions ($\omega=0$, $\homega=\frac1n$, denoted $\td M$). The net
effective potential from equation~(\ref{eq:veffst}) is
\bea
V_p(\bar{a},\varphi,\psi)&=& \bar{a}^{-3}\left[\rho_{\sss\td
  N}e^{(p-\frac{\nc}{2})\sqrt{B}\psi}e^{(p-\frac \nc2)\sqrt{\frac
    A2}\varphi}+\rho_{\sss\td M}e^{-(1+\frac{\nc}{2})\sqrt{B}\psi}
   e^{-(1+\frac\nc2)\sqrt{\frac A2}\varphi}\right] \label{eq:veffb3}
\eea
This scenario is similar  to those analyzed
in \cite{abe,branwat_isot,arakay,kaya_vol,bercli}. As we will now explore,
the result of
including a gas of branes is the improved stability of the radion.
Inspection of~(\ref{eq:veffb3}) reveals
that provided $p>\frac\nc2$, all internal directions will be
stabilized, since there are both rising and falling exponentials depending
on $\psi$:   
\bea
V_p(\bar a,\psi)&=&\rho_{\td N}e^{(p-
\frac{\nc}{2})\sqrt{B}\psi}\bar{a}^{-3}+\rho_{\td M}e^{-(1+\frac{\nc}{2})\sqrt{B}\psi}\bar{a}^{-3}\ .\label{eq:veffp}
\eea
where again we have assumed $\varphi=0$.
This has a nontrivial minimum close to $\psi=0$ provided that
 $p>\frac\nc2$. Since string theory requires $\nc=6$, the
presence of ($p>3$)-branes in the compact directions will stabilize the
moduli.  
\par
However a more detailed analysis may be necessary to realize these stability
conditions: According to the heuristic argument of Alexander, Brandenberger,
and Easson~\cite{abe}, winding modes will generically intersect in $2p+1$
dimensions, so that only objects with $p\le2$ should remain wound in the
6~compact directions. In this case, the stability requirement will not be
satisfied. On the other hand, a quantitative investigation 
should account for the larger phase space
once the $\bar a$ directions have grown large, thus decreasing the probability
of annihilation, and perhaps leaving some extended objects with $p>\frac\nc2$.
As well, such an analysis should be carried out within the product space
$\mathcal{T}_4\times\mathcal{T}_2\times\mathcal{T}_3$ argued for by~\cite{abe}, not
the $\mathcal{T}_n\times\mathcal{T}_3$ topology we have
considered. We do note though that, as long as the shape moduli is frozen, the exact shape of the tori does not matter and our stability analysis
for the volume  still applies.
\par 

\section{No-Go Result in Type II String Theory}  
\label{sec3}
In the previous section we saw that the radion can be stabilized by a brane gas
when the dilaton is assumed to be fixed; similarly one can show the corresponding
result when the roles of the dilaton and radion are interchanged.  
One may naturally wonder
whether both of these moduli  can  be simultaneously stabilized using the most
general combination of string and brane sources. As we now show, this is
impossible to do with the conventional (winding, momentum or oscillator) string
and brane excitations.  The argument is made in two steps, starting first with
gases where each string or brane has only one kind of excitation  (``simple states''),
although different species of strings or branes are allowed to co-exist.  We then
extend the argument to the more general case where individual components of the gas
have more than one kind of excitation (``mixed states''). 

\subsection{``Simple States''}

We consider the situation when the strings/branes have 
nontrivial wrapping  of only some extra dimensions, i.e. it doesn't wrap the large dimensions. 
They thus appear point-like to 4D observers and 
redshift like nonrelativistic dust,
$\bar{a}^{-3}$, corresponding to 
$d=3$, $\omega=0$ in (\ref{eq:veffst}). 
The equations of motion for the radion and the dilaton
in the presence of such sources are
\be
\ddot{\varphi}+3H\dot{\varphi}=-{\p V_{\rm{eff}}(\varphi,\psi)\over \p \varphi}
\ee
\be
\ddot{\psi}+3H\dot{\psi}=-{\p V_{\rm{eff}}(\varphi,\psi)\over \p \psi}
\ee
with
\be
V_{\rm{eff}}(\varphi,\psi)=\bar{a}^{-3}\sum_i \rho_ie^{2\mu_i\varphi+2\nu_i\psi}
\label{sum}
\ee
where the sum runs over all possible string and brane states. As
summarized in Appendix~\ref{app:exps}, the exponents $\mu_i,\nu_i$
depend only on the effective equation of state parameter $\omo$ along
the extra dimensions (\ref{eq:veffst}), and the coupling exponent
$\al$ of these states to the dilaton in the string frame:
\be
S_{{\rm gas},i}=-\int d^{\,D}x\ \sqrt{-{g}}\,e^{-\al_i\phi}\,\rho_i\,
b^{-\nc(1+\omo_i)}\,{a}^{-3}\ .
\label{gen-action}
\ee
where $\al=0$, for the fundamental strings since the Nambu-Goto action does
not contain any dilaton coupling, while for branes $\al=1$, originating from
the dilaton coupling in the DBI action. 

After performing the conformal transformations involving the radion
and the dilaton (see Appendix~\ref{app:dimred}) the above action
gives rise to the effective potential (\ref{eq:veffst}) in the 4D
Einstein frame with 
\be
\mu_i={1\over 2\sqrt{\nc+2}}\left[1-\al_i-\nc(\omo_i+{\al_i\over 2})\right];\quad
\nu_i=-\sqrt{\nc\over2(\nc+2)}\left(\omo_i+{1\over 2}\right)
\label{gen-munu}
\ee

As noted earlier, the value of $\omo$ depends upon whether
 the mode in question has winding, momentum or string oscillations.
To analyze the stability of a potential which is a sum over such modes,
we will use the technique of \cite{prashanth}: we identify the directions in the
$\varphi$-$\psi$ plane in which there exists a rising exponential contribution.
If such directions are sufficiently numerous, the system is completely
stabilized.
An exponential of the form  $\rho_i e^{2\mu_i\varphi+2\nu_i\psi}$ rises
most steeply along the direction $\cos\te_i\hat{\varphi}+\sin\te_i\hat{\psi}$
where $\tan\theta_i = \nu_i/\mu_i$.
In the range 
\be
\te=(\te_i-\pi/2,\te_i+\pi/2)
\label{range}
\ee
 there is a rising potential (wall)
while along the other half-plane the potential asymptotically falls to zero.
Since our potential is a sum of exponentials, it is clear that: \\
(I) There can be at most a single local minimum and no local maxima.\\
(II) Such a minimum exists only if the potential grows in any direction away
from the minimum.  Thus there must exist angles $\theta_i$ for which the
ranges of angles in (\ref{range}) cover the entire plane.
 
By looking at the different directions of steepest ascent of the
exponentials it is easy to verify whether (II) is satisfied.
Curiously, for brane sources ($\al_i=1$) we find a result that is
specific to $d=3$ large  dimensions: the direction of steepest ascent
is the same (modulo $\pi$) for
winding, momentum or any other  modes.  For the winding modes,
it is given by
\be
\tan\te=\frac{\nu_i}{\mu_i}=\sqrt{{2\over\nc}}\quad\Ra\quad \te={\pi\over
  6} +\left\{\begin{array}{cc} 0, & p>3 \\ \pi, & p<3
\end{array} \right.
\label{eq:bangle}
\ee 
while for momentum modes 
\be
\te = {\pi\over 6}
\ee
 for all $p$,
as illustrated in figure~\ref{steepdirs}.
Thus using, say, a gas of 2- and 6-branes in type IIA theory, or a
gas of 1 and 5-branes in IIB theory, one could stabilize all 
directions of the
$\varphi$-$\psi$ plane except for those orthogonal to the
direction of steepest ascents, that is, $\te={\pi\over 6}\pm
{\pi\over 2}$ (see figure).  Along these directions the potential
is flat, and there is a zero mode.

\EPSFIGURE[ht]{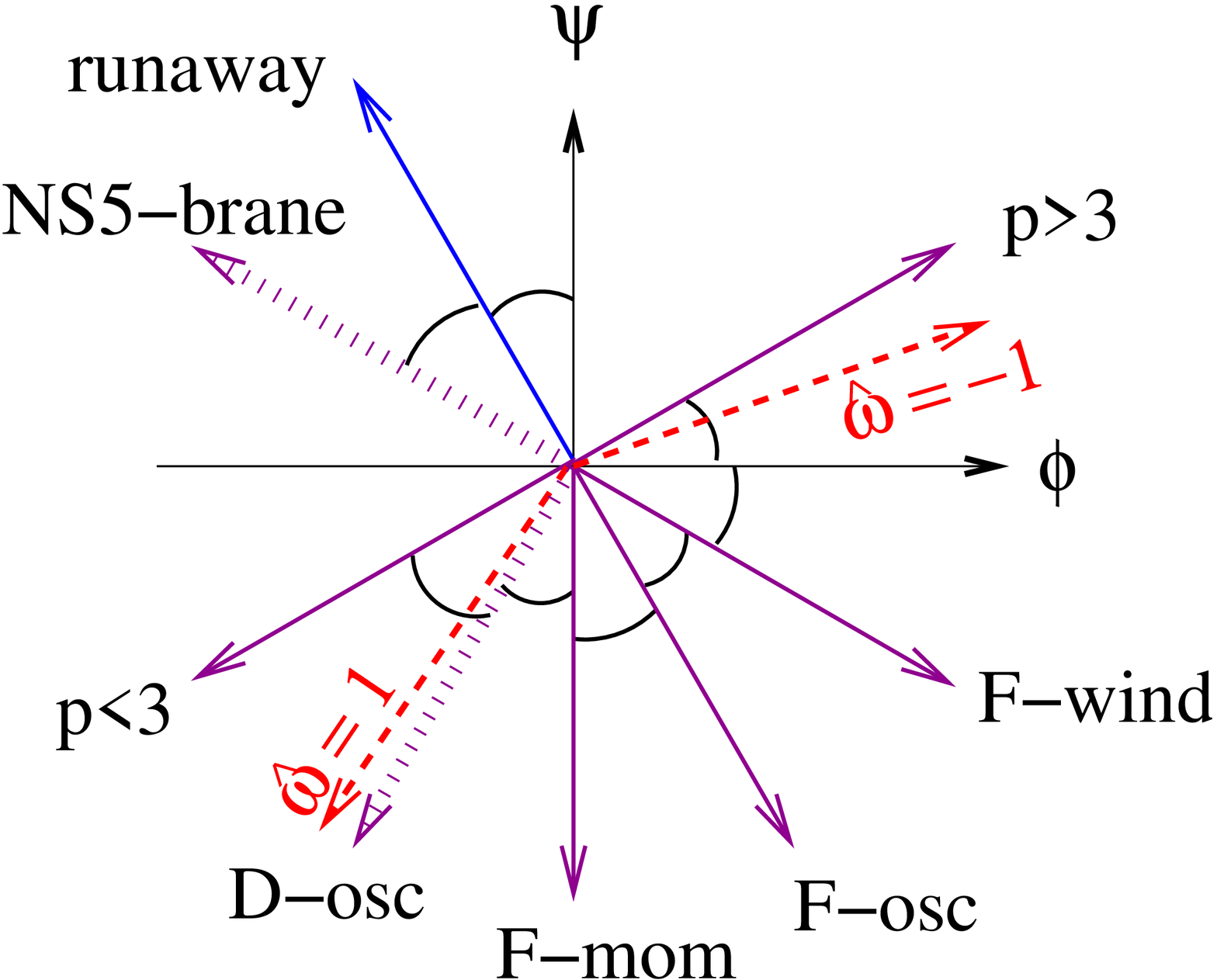,width=3.75in}{Directions of steepest ascent 
in the $\phi$-$\psi$ plane for
contributions from different brane gas sources, described in the text.
Long dashed lines are for the hypothetical F-string sources with $\omo=\pm1$.
All marked angles are 30$^\circ$.  Short dashed lines are for the sources
described in section \ref{exo}.
\label{steepdirs}}

To lift the flat direction, one has to incorporate  fundamental
string sources, with $\al_i=0$. According to Table \ref{tab:exps}
for F-strings, the angle 
\bea
\tan\te=\frac{\nu_i}{\mu_i}=\sqrt{\frac2\nc}\,
\frac{\homega_i+\frac12}{\homega_i+\frac1\nc}
\eea
depends on $\omo_i$, and so the analysis has to be done separately
for each case. For 
the string momentum, oscillatory and winding modes,
$\omo=1/\nc,\ 0$ and $-1/\nc$\ respectively. From
(\ref{gen-munu})  one then finds
\be
\te_{\rm{mom}}=-{\pi\over 2},\quad \te_{\rm{osc}}=-{\pi\over 3},
\quad \te_{\rm{wind}}=-{\pi\over 6}
\ee
All of these modes provide an ascending potential in the direction of
$\te=-{\pi\over 3}$, which is the flat direction when only brane
sources are present,  but not along the opposite direction
(see figure~\ref{steepdirs}). Thus after including
the string modes, there is no longer a zero mode, but there is a
runaway direction along  $\te_{\rm{run}}={2\pi\over 3}$.

In fact we can make an even stronger statement.  Suppose that there
are other string sources we may be unaware of; nevertheless
their equations of state should satisfy the weak energy condition
$-1\le\omo\le 1$.  
Using the Table 1 entries for general string sources and 
varying $\omo$ over this range 
gives an angle of steepest ascent in the range
\be
\pi-\tan^{-1}{3\sqrt{6}\over 5}\le
\te\le \tan^{-1}{\sqrt{6}\over7} \quad\ra\quad
-124^\circ\lsim \te \lsim 19^\circ
\ee
which again fails to lift the $\te_{\rm{run}}$ direction,
as shown in figure~\ref{steepdirs}.
\par
An alternative understanding of the moduli instability can be
directly inferred by a reparametrization of the effective
potential~(\ref{eq:veffst}) in terms of new fields
$\chi=\sqrt{B}\psi+\sqrt{\frac A2}\varphi$ and
$\eta=\sqrt{\frac A2}\psi-\sqrt{B}\varphi$. The result for 
an arbitrary string and brane gas with three large directions and
$\omega=0$ is
\bea
V_{{\rm eff},i}(\alpha=0)&=&\rho_i\,
e^{\lb-\nc(\homega+\frac14)+\frac12\rb\chi}\,
e^{-(\frac\nc2+1)\frac\eta2}\,\bar{a}^{-3}\hbox{\ (string)}\label{eq:stf}\\
V_{{\rm eff},i}(\alpha=1)&=&\rho_i\,
e^{-\nc(\homega_i+\frac12)\chi}\,\bar{a}^{-3}\qquad\qquad\quad\ \,\hbox{\ (brane)}\ 
\label{eq:bf}
\eea
Through a combination of sources it is possible to 
to stabilize the $\chi$
mode; however string sources will only cause $\eta$ to grow, and
a brane gas does not couple to $\eta$.  Thus $\hat\eta = \sqrt{\frac A2}\hat\psi 
- \sqrt{B} \hat\varphi$ is the unstable
direction in field space, in terms of the unit vectors 
$\hat\varphi$, $\hat\psi$.  This direction corresponds to the line
\bea
\psi=-\sqrt{\frac{A}{2B}}\,\varphi=-\sqrt{\frac \nc2}\,\varphi\ ,
\eea
which coincides precisely with the  principal runaway
direction identified previously.

\par
The above result is consistent with ref.\ \cite{bercli}, which used a
perturbation analysis in the BGC scenario to show that the inclusion
of branes alone is not enough to stabilize both the dilaton and
moduli fields. Thus some other potential is needed to stabilize one
of the moduli. Given such a potential, BGC
does provide a mechanism of stabilizing the other degree of freedom
provided that branes with $p>\frac\nc2$ are
present. 
\par
The factorization of the effective potential is a coincidence of 
having  $d=3$ large dimensions, as can be seen from the nontrivial
dependence on $d$ in eq.\ (\ref{eq:veffst}).  For scenarios other
than $d=3$, the gas of strings and branes is able to
stabilize both fields. 

Finally, we note that in the above analysis we did not consider branes or
strings which wrap some of the large three dimensions ($\om<0$); these do not
give any additional leverage for stabilizing the radion.

\subsection{``Mixed States'' and Massless Modes}
So far we have only considered states which are purely oscillatory,
winding or momentum modes.  More generally, strings could 
have a combination of such excitations.
Can such mixed modes help in stabilizing the moduli? The answer,  unfortunately is no. For massive modes the reasons are similar to the case of the simple states. The massless modes\footnote{ Although, in the type II theory
  these are excitations of the tachyon which
are removed by the GSO projection, such massless winding states are
allowed  in  the 
heterotic string theory. } have to be analyzed separately but they do not alter the conclusions. 

In the string frame, the string spectrum is
\be
m_F^2={m^2\over b^2}+N_{\rm{osc}}M_s^2+w^2b^2M_s^4
\label{spect}
\ee 
where $M_s$ is the string scale and the three terms on the right hand
side correspond to the momentum, oscillatory and winding pieces
respectively. The source action for the strings (\ref{gen-action})
then becomes 
\be
S_{\rm{str}}=\int d^Dx\ \sqrt{-{g}}\,a^{-3}\,b^{-\nc}\,\sqrt{m^2_F(b)+p^2}
\ee
where $p$ is the momentum along the non-compact directions. After performing the dimensional reduction and conformal redefinitions, as usual we find that the  effective potential for the canonical radion and dilaton coming from a gas of such states is given by
\be
V_{\mt{eff}}(\varphi,\psi)=\rho=nE(\varphi,\psi)
\ee
where $n$ is the number density, and $E(\psi,\phi)$ is the energy of these states which  depends on both the dilaton and the radion.  Since we already know how the exponents look like for individual momentum, winding and oscillatory modes, it is easy to see that for the more general case the energy is just given by
\be
E(\psi,\varphi)=\sqrt{M_s^2[m^2e^{4(\mu_m\varphi+\nu_m\psi)}+w^2e^{4(\mu_w\varphi+\nu_w\psi)}+N_{osc}e^{4(\mu_o\varphi+\nu_o\psi)}]+p^2}\equiv\sqrt{m_F^2+p^2}
\ee
with
$$\mu_m=0\ ;\ \mu_o={1\over2\sqrt{8}}\mx{ and } \mu_w={1\over\sqrt{8}}$$
and
\be
\nu_m=-{1\over\sqrt{6}}\ ;\ \nu_o={\sqrt{3}\over2\sqrt{8}}\mx{ and } \nu_w=-{1\over2\sqrt{6}}
\ee

First let us focus on the massive modes, for which one can ignore the momentum $p$.
The key observation is that since  $N_{\rm{osc}}\ge 0$ for massive modes, the
effective potential obtained in the Einstein frame must still satisfy
conditions (I) and (II) above in order to have a local minimum. Since again such  potentials can have at most one
minimum and no maximum, if there exists a minimum, the potential has
to keep rising along any direction as one tends towards infinity.
Thus to determine whether there is a minimum, it suffices to
investigate the behaviour at infinity in the $\varphi$-$\psi$ plane.
Going far enough toward $\infty$ along a generic direction, one of the
three terms in (\ref{spect}) will dominate, and then our  previous
analysis applies, which assumed the presence of only one term in a
given source.  In the special direction where $b$ remains  constant,
no one term dominates, but they all remain proportional to each
other, behaving like a single term,  so again the previous analysis
remains valid. 

Next let us focus on the massless modes, for example, the ones considered in \cite{subodh}. In this case 
$N_{\rm{osc}} < 0$;  and depending on the winding and momentum quantum numbers one could have $m_F^2 \sim (m/b-wbM_s^2)^2$
leading to a minimum at $b^2=(m/w)M_s^{-2}$ (which is the self-dual
radius when $m/w=1$) \cite{subodh}.  For these modes one can easily verify  that the mass function  can be cast as
\be
m^2_F(\varphi,\psi)\sim e^{\sqrt{2}\varphi'}\left(e^{{\psi'\over \sqrt{6}}}-e^{-{\psi'\over \sqrt{6}}}\right)^2
\label{massless}
\ee
where
\be
\psi'={\sqrt{3}\over 2}\varphi +{1\over 2} \psi
\ee
(as one can find by carefully tracing back the conformal transformations) is really the string frame radion and
\be
\varphi'={1\over 2}\varphi -{\sqrt{3}\over 2} \psi
\ee
is the orthogonal direction.
As one can see, the mass (\ref{massless}) and  the potential have a minimum at $\psi'=0$
and hence the massless states  stabilize the $\psi'$ direction,  as argued in
\cite{subodh}.  However, $\psi'$ also precisely coincides with  the direction that could be
fixed just with winding branes. Thus we are still left  with the orthogonal runaway
direction ($\varphi'\ra -\infty$) that we found earlier.

\subsection{Exotic States}
\label{exo}
We have seen so far that ordinary D-brane and string states are unable to stabilize both
the radion and the dilaton simultaneously. 
We now briefly discuss how stabilization might be achieved using some less
conventional kinds of branes.

One kind of exotic state which has been considered  \cite{subodh} are  quantized  D-string
modes.  Whether it is justified to derive these from 
the Nambu-Goto action like for F-strings seems doubtful, since the D-string is a solitonic
object, but for completeness we have derived the exponents corresponding to the different 
D-string modes\footnote{ See Appendix \ref{app:dosc} for details.}.  Although oscillator
excitations do provide a new direction in field space whose potential has a steep direction,
this direction overlaps with ones from other more conventional sources, and  do not  affect
our no-go result.  The direction of steepest ascent for  the D-string oscillator modes is
derived in Appendix \ref{app:dosc},  and is shown in figure~\ref{steepdirs}.   However if
massless D-string modes are also allowed in the string theory spectrum, they can lift the
runaway direction in conjunction with other modes, as has been argued in \cite{subodh}.

Another possible source that provides an effective potential is the NS5-brane.\footnote{We
thank Ali Kaya for pointing this out to us.}\  Its tension
behaves as $T_5^F\propto g_s^{-2}$, so an NS5-brane wrapping the internal
manifold corresponds to $\alpha=2,\ \homega=\frac{-5}{n},\ \omega=0$; this results in the
coupling coefficients
\be
\mu_5^F=-(\frac n2-2)\frac{1}{\sqrt{n+2}}\mx{ and } \nu_5^F=(10-n)\sqrt{\frac{2}{n(n+2)}}
\ee
Thus, for the case $n=6$, the potential rises maximally in the direction
$\psi=-\frac{1}{\sqrt{3}}\phi$, corresponding to the angle $\theta_{F5}=150^\circ$ in
figure~\ref{steepdirs}. This does not coincide with the runaway direction identified in
figure~(\ref{steepdirs}), but is close enough so that NS5-branes
in conjunction with strings or D-branes can stabilize all the moduli. 

\section{Adding Superpotentials}
\label{sec4}
Since it is not possible to fully stabilize all moduli using the
D-brane/string gas, we investigate the dynamics of a system in which
the brane gas is present simultaneously with an external
stabilization mechanism.  A typical potential which could arise in
string-motivated supergravity theories is the one which is generated
by gaugino condensation in an SU(N)  gauge sector.  Although it might
seem redundant to consider partial modulus stabilization by a brane
gas when there is already a potential at zero density, there could
actually  be several  benefits: for example, the brane gas can
prevent the problem of the moduli overshooting the desired minimum
\cite{brustein}, as we investigate in this section. 

\subsection{Gaugino Condensation Potential}
We briefly review the derivation of the nonperturbative gaugino
condensate potential
in low-energy effective supergravity, starting 
with the 10 dimensional spacetime which is assumed to be a product of 4D
noncompact external spacetime and a 6D compact internal manifold. 
We limit our present discussion to the dynamics of the radion, $\psi(x)$.
A similar discussion should apply for 
more than one moduli field, but for simplicity  we
assume that all other moduli ({\it i.e.}, complex structure and dilaton) 
have been stabilized.
The radion appears in the full metric as   
\be
{d} s^2=g_{\mu\nu}dx^\mu dx^{\nu}+e^{2\psi}g_{\ms\ns}dy^{\ms}dy^{\ns}
\label{metric}
\ee 
In supergravity, the radion is the real part of a chiral field,
\be
T=X+iY\equiv e^{4\psi} + iY
\ee
Dimensional
reduction of the supergravity action yields an effective four dimensional
theory of gravity coupled to the complex scalar field $T(x)$
\be
S_{\rm{}}=M_p^2\int d^{\,4} x \sqrt{-g}
\left[\,{R\over 2} +{\cal K}_{T\bar{T}}\,\p_\mu T\, \p^\mu\bar{T} 
-e^{\cal K}({\cal K}^{T\bar{T}}D_{T}{\cal W}\,
\overline{D_{T}{\cal W}}-3|{\cal W}|^2)\,\right]
\ee
where  ${\cal K}(T,\bar{T})$ and  ${\cal W}(T,\bar{T})$ are the
K\"ahler potential and superpotential respectively, while $K_{T\bar{T}}$
is the K\"ahler metric given by
\be
{\cal K}_{T\bar{T}}=\frac{\p^2{\cal K}}{\p T\p\bar{T}}
\ee
We have also performed a conformal transformation of the
four dimensional metric:
\be
g_{\mu\nu}\to e^{\Ds\psi}g_{\mu\nu}=e^{6\psi}g_{\mu\nu}
\label{rad-conf}
\ee

The kinetic and potential terms for $T$ are computed from 
${\cal K}(T,\bar{T})$ and  ${\cal W}(T,\bar{T})$, where the K\"ahler
potential for $T$ is ${\cal K}=-3\ln [T+\bar{T}]$
while as in \cite{kklt} we use the superpotential
\be
{\cal W}={\cal W}_0 +Ae^{-aT}
\ee
which would be obtained through gaugino condensation in a theory
with a simple gauge group. For instance, for $SU(N)$,
$a=2\pi/N$. The constant term $W_0$ represents the
effective superpotential due to any fields that have
been fixed already \cite{GKP}, such as the dilaton and complex structure
moduli.\footnote{Although recent authors have pointed out that a
  proper construction of the KKLT mechanism includes other non-perturbative
  contributions to the K\"ahler potential, we are primarily concerned
  with addressing the overshoot problem which still exists despite
  their inclusion~\cite{deAlwis,quevedo}. That is, the mechanism proposed
  still provides an attractor solution despite changes to the form of the K\"ahler potential.}

The scalar-tensor action  then reads
\be
S_{\rm{}}=M_p^2\int d^4 x \sqrt{-g}\left[{R\over 2} +K -V\right]
\ee
where the kinetic ($K$) and potential ($V$) terms are given 
by 
$$
K=-3\frac{\p_\mu T\,\p^\mu\!\bar{T}}{|T+\bar{T}|^2}=
-12\,\p_\mu\psi\,\p^\mu\psi -{3\over 4}{e^{-8\psi}}\,\p_\mu Y\p^\mu Y
$$  
and 
\be
V={E\over X^{\al}}+{1\over 6
X^2}\left[aA^2(aX+3)e^{-2aX}+3W_0Aae^{-aX}\cos(aY)\right]
\label{rad-potential}
\ee
To arrive at the potential (\ref{rad-potential})) we have also included the
potential energy coming from an anti-D3 brane (first term) as in \cite{kklt},
which is needed in order to have a nonnegative vacuum energy density. The
coefficient $E$ is a function of the tension of the brane $T_3$ and of the warp
factor, if there are warped throats \cite{KS} on the Calabi-Yau manifold. The exponent
$\alpha$ is either $\alpha=2$ if the anti-D3 branes are sitting at the end of a
warped throat.  Otherwise $\alpha=3$ corresponding to the unwarped region. If a
warped region exists, it is energetically preferred.

The imaginary part of the K\"ahler modulus, the axion $Y$,
has stable minima at $Y=(2n+1)\pi/a$ (assuming $W_0Aa>0$).  We will
integrate this field out and focus on the dynamics of the radion,
whose kinetic term is\footnote{The kinetic part of the action for the radion 
can also be derived directly from the Einstein-Hilbert action
${S_{10}}=\int d^{10}x\sqrt{-\hat{g}}\hat{R}$
contained in the full 10D supergravity action. 
Using a consistent dimensional reduction ansatz for the metric of the form 
(\ref{metric}) one obtains
${S}_{4}=\int d^4x\sqrt{-g}e^{\Ds\psi}[R-\Ds(\Ds-1)(\p\psi)^2+\dots]$.
A conformal transformation precisely of the form
(\ref{rad-conf}) is needed to convert the action $S_4$ to
Einstein frame, from which one recovers the kinetic piece of the radion given
here.
} $12M_p^2(\p \psi)^2$, and whose potential becomes
\be
V=Ee^{-4\al
\psi}-\frac12\left(W_0Aae^{-aX}-aA^2e^{-2aX}\right)e^{-8\psi}+\frac16
a^2A^2e^{-2aX}e^{-4\psi}
\ee
It is convenient to rescale $\psi\to\psi/\sqrt{24}$
so that the kinetic term is canonically normalized. 
The action becomes
$$
S_{\rm{}}=M_p^2\int d^4 x \sqrt{-g}\left[{R\over 2} -{(\p
\psi)^2\over 2}\phantom{\frac16}\right.
\qquad\qquad\qquad\qquad\qquad\qquad\qquad\qquad\qquad\qquad\phantom{.}$$
\be
\left. 
-\left(Ee^{-2\al_1\psi}-\2\left(W_0Aae^{-aX}-aA^2e^{-2aX}\right)e^{-2\al_2\psi}+\frac16
a^2A^2e^{-2aX}e^{-2\al_3\psi}\right)\right] 
\label{finalaction}
\ee
where
\be
\al_1={\al\over\sqrt{6}},\ \ \al_2={2\over\sqrt{6}},\ \ 
\al_3={1\over\sqrt{6}}, \quad X=e^{2\al_3\psi}
\ee
The potential has three distinct regions (see figure \ref{potential}, solid
curve). (1) For $\psi$ large and negative $V(\psi)$ is dominated by the
antibrane contribution, if $\alpha=3$, or by a combination of the antibrane
term and the term proportional to $e^{-2\al_2\psi}$ if $\alpha=2$.  In either
case, the potential is to a good approximation a pure exponential in $\psi$,
which will be relevant for the analytic solutions we discuss in the next
subsection. (2) For $\psi\sim 1$,
the different terms in the potential are comparable, creating a minimum
at $\psi_{\rm min}$, followed by a potential barrier at $\psi_{\rm max}$. 
(3) For $\psi\gg 1$  the antibrane term again dominates, since the other terms are
exponentially suppressed by $e^{-aX}$. 

\EPSFIGURE[b]{pot2.eps,width=0.5\hsize}{Radion potential with vacuum gaugino condensate
potential (solid line) and potential at nonzero brane gas density.\label{potential}}

\subsection{Attractor solution with brane gases} 

We now consider the effect of augmenting the vacuum potential in 
(\ref{finalaction}) with
the contribution from a brane gas.  Without the brane gas, 
the dynamics of the radion depend sensitively on the initial conditions. 
If we start with $\psi>\psi_{\rm max}$ (the position of the bump in the 
potential), $\psi$ runs to infinity, where the extra dimensions are
decompactified.\newline Generically one might expect the radion to start closer
to the Planck size with $\psi < 0$, so that there is a possibility of reaching
the stable minimum at $\psi = \psi_{\rm max}$.

Since the vacuum potential in the region $\psi < 0$ is well approximated by an
exponential,
the radion quickly reaches the attractor solution discussed in \cite{anupam2};
it tracks the minimum formed between the exponential potential and the
rising part of the ``brane-gas potential,''  shown as the dashed line
in figure \ref{potential}.   This attractor behavior 
washes out the effect of initial conditions.  As long as the attractor
is reached before the field has passed the position of the minimum,
this will allow $\psi$ to settle into the minimum and avoid the
overshoot problem.

Let us recapitulate the details of the attractor solution.
The rising part of the brane-gas potential originates from the winding modes of
$p$-branes with $p>3$. In this region the Friedmann equation and the
equation of motion for $\psi$ read (in $M_p=1$ units)
\be
H^2\cong\3\left({\frac12\dot{\psi}^2}+Ee^{-2\al_1\psi}+\rho_{p}e^{2\nu_p\psi}\right)
\ee 
{ with } $\rho_{p}=\rho_{p}^0\left({a\over a_0}\right)^{-3}$ and
\be
\ddot{\psi}+3H\dot{\psi}\cong 2\left(\al_1 Ee^{-2\al_1\psi}+\nu_p\rho_{p}e^{2\nu_p\psi}\right)
\ee
respectively. The exponents $\nu_p$ for $p$-branes' coupling to the 
canonically normalized radion were derived in the previous section,
\be
2\nu_p=\sqrt{\frac{2}{\nc(\nc+2)}}\left(p-{\nc\over
2}\right)=\sqrt{\frac{1}{24}}\left(p-3\right)
\ee
where $n=6$ is the number of extra dimensions.
This kind of system was studied in \cite{biswas}, where it 
was shown that there exist tracking solutions in which 
the energy of the scalar field tracks that of the branes: 
\be
e^{\psi}=\left[{E\over\rho_p}\left({\al_1(\al_1+\nu_p)-3/4\over 
\nu_p(\nu_p+\al_1)+3/8}\right)\right]^{1\over
2(\al_1+\nu_p)}\equiv\left[{E\over\rho_p}r\right]^{1\over 2(\al_1+\nu_p)} 
\label{twofluidsol}
\ee
This relation implies that both the potential  
and kinetic energy of the radion remain
proportional to the energy density of the branes,
\be
V(\psi)= r^{-1}{\rho}_p\,e^{2\nu_p\psi}={8(\nu_p+\al_1)^2-3\over 3(1+r)}K
\label{proportions}
\ee
The steepest brane-induced effective potential occurs for the maximal value of
$p$, $p=6$; this
provides the greatest resistance to expansion of the internal manifold
and will be the most effective case for avoiding the
overshoot problem. Eq.\ (\ref{proportions}) also shows that for $p=6$ 
the ratio of kinetic to potential energy is minimized. 
For example, if $\al=2$ and $p=6$ we
have $\al_1=\sqrt{2/3}$, $2\nu_6=\sqrt{3/8}$,  leading
to ${K /V}={12/23}$.

In passing we note that such tracking solutions correspond to a power law
expansion of the universe
\be
a(t)=a_0\left({t\over t_0}\right)^{(2/3)(1+\nu_p/\al_1)}=
a_0\left({t\over t_0}\right)^{11/12}
\label{power}
\ee
The universe does not accelerate during this phase.  However, as was
found in \cite{biswas}, when
the analysis is carried out including the dilaton, acceleration can
be obtained.
\subsection{Addressing the Overshoot Problem}

The above discussion implies that the overshoot problem will be avoided in
the presence of a brane gas so long as the attractor solution can be reached.
This means that for a given initial value of $\psi$, the initial energy
density in the brane gas, $\rho_p e^{2\nu_p\psi}$, must be sufficiently large.
If not, the brane density is diluted too quickly by the expansion
of the universe and the system evolves according to the vacuum potential.

We have confirmed these expectations by numerically integrating the coupled
system of Friedmann and radion equations, which we 
 illustrate with a specific example.  In the potential
(\ref{rad-potential}) we consider an antibrane in a warped throat,
with $\alpha=2$.  Its tension is tuned to give a Minkowski minimum
as shown in figure \ref{potential}, which illustrates the case where
$E = 0.00889,$ $a = 2.1,$ $A = 0.9,$ and $W_0 = 0.25$.  We first verified that
indeed this potential suffers from an overshoot problem, shown in figure
\ref{overshoot}.  Starting from an initial condition $\psi \lsim - 0.17$,
the field runs away to $\infty$.

Interestingly, overshooting can be prevented by initial brane densities
which are many orders of magnitude smaller than the initial potential 
energy of the radion.  Figure \ref{undershoot} shows the evolution
starting from exponentially large initial radion potential energy,
with $\psi_0 = -100$ and $p=6$, for several initial brane densities, parametrized by
$\zeta = \rho_p e^{2\nu_p\psi}/V_0(\psi_0)$, where $V_0$ is the potential
of the radion alone, excluding the brane gas contribution.   The result 
shows that even for initial brane gas energy densities which are only
$10^{-18}$ $V_0(\psi_0)$, overshoot can be prevented.  For different initial
values, the exponent $\log_{10}(\zeta)$ scales linearly with $\psi_0$.  This
behavior can be understood analytically, as shown in Appendix D.
The minimum required value of $\zeta$ is given by
\be
\log_{10}\zeta \approx -0.43\left[2\al_1\psi_0\left(1-{3-4\nu_p\al_1\over 4\al_1^2}\right)\right]
\ee
The intuitive explanation for this result is that the radion energy
initially falls more quickly than that of the brane gas.  What counts is not the
initial ratio of brane gas to potential energy; rather it is the ratio at the time when
$\psi$ is close to its nontrivial minimum.   This mechanism has been pointed out in
\cite{dealwis} (see also \cite{ovrut,sorbo})
as a generic way of solving the overshoot problem, using general sources
of energy density.  Brane gas cosmology provides a concrete setting where this idea can
be used advantageously.

\DOUBLEFIGURE[ht]{solns.eps,width=\hsize}{solns-brane2.eps,width=\hsize}
{Evolution of $\psi$ without brane gas, for several initial values $\psi_0 =
-0.15$, $-0.16$, $\cdots, -0.2$, illustrating overshoot.\label{overshoot}}
{Solutions with $\psi_0=-100$ and different initial densities of brane gas,
near the borderline of overshooting.\label{undershoot}}

When the modulus has reached its stable minimum, we are still left
with a gas of branes, whose energy density is comparable to the
energy density in the scalar fields; otherwise the brane gas would
not be effective in slowing the rolling of the modulus.  At the
bottom of its potential, the scalar field oscillates and and its
energy density redshifts as $a^{-3}$ just like the brane gas.  The
result is a matter dominated universe.  We must assume that inflation
begins  some time after this in order to dilute the branes and reheat
the universe.  Work on smoothly connecting the modulus stabilization
with the beginning of inflation is in progress.

\section{Conclusions} 
\label{sec6} 

In this paper we used dimensional
reduction to derive the  effective action for a gas of strings and
$p$-branes, giving a contribution to the effective  potential for the
radion and dilaton.  In a gas of strings only, this potential 
could stabilize the radion provided there was only one extra
dimension, but not the dilaton.
dilaton. Including $p$-branes allows for the stabilization 
of either the dilaton or radion if
$p>\frac d2$.  However, the brane gas is insufficient
for stabilizing both moduli simultaneously, for the type II strings
we consider, which have no massless winding modes. Rather, only a
linear combination of the moduli can be stabilized by the brane
gas.

It thus seems likely that external potentials are needed for
modulus stabilization.  However the brane gas can still play an
interesting role in helping the moduli settle into their typically
shallow minima, avoiding the overshoot problem.  An attractive feature
of this mechanism is that the brane gas can initially be many orders
of magnitude smaller in energy density than the potential energy of
the moduli and still be effective in slowing the rolling of the
moduli, since the brane gas energy redshifts more slowly.  There
is therefore no need for finely-tuned initial conditions.

In this work we have ignored quantum corrections, as well as
higher-derivative corrections to the dilaton gravity action.  The
first approximation is justified for weak   string coupling,
$g_s=e^{\phi}<<1$. In this regard, the runaway direction found
in Section~\ref{sec:BGC} corresponds to $\phi\rightarrow-\infty$,
showing that quantum corrections cannot lift this flat direction at
large field values.  Of course it is possible that such corrections
could lead to a metastable minimum along the flat direction, which
would be a loophole in our no-go result.

\section*{Acknowledgments}
We would like to thank Robert Brandenberger,  Ali Kaya, Anupam Mazumdar, Subodh Patil,
and Horace Stoica for useful
discussions. This work was supported in part by NSERC and FQRNT.

\section*{Note Added} As this work was being finished, similar
results were given in \cite{Easson:2005ug}.  Our results were
presented the week before, at the McGill Workshop on String Gas
Cosmology, 30 April 2005.

\appendix
\section{Dimensional Reduction}
\label{app:dimred}
We provide here a brief review of the standard dimensional-reduction
procedure, following the procedure of~\cite{batwat,cghw}. Our starting
point is $D$-dimensional dilaton-gravity together with a generic
contribution of gas. This system is described by
\bea
\label{ap:dg}
S_{II}&=&\frac{1}{2\kappa^2}\int d^Dx\sqrt{-G}e^{-2\phi}\lb
R+4G^{MN}\nabla_M\phi\nabla_N\phi-\frac{1}{12}H_{\mu\nu\alpha}H^{\mu\nu\alpha}\rb\\
S_m&=&\int d^Dx\sqrt{-G}e^{-\alpha\phi}\rho\ ,
\hspace{2em}\rho=\sum_i \rho_ia^{-d(1+\omega_i)}b^{-\nc(1+\homega_i)}\ ,\label{ap:mat}
\eea
for some initial density $\rho_i$. The dimensional reduction procedure will
focus on the string action~(\ref{ap:dg}); but, by tracking the
transformation rules, we can later also reduce the matter components. We obtain an effective
theory of BGC by first transforming the string action~(\ref{ap:dg}) to the Einstein frame through the
 conformal transformations~\cite{birdav}

\begin{minipage}{\textwidth}
\bea
G_{MN}\rightarrow \td G_{MN}&=&\Omega^2G_{MN},\ \Omega=e^{-A\phi}\ ,\
A=\frac{2}{D-2}\nonumber\\
R\rightarrow\td R: R&=&e^{-2A\phi}\td R-2(D-1)e^{-A\phi}\lb
e^{-A\phi}\rb_{;MN}\td G^{MN}\nonumber\\&&-(D-1)(D-4)\lb e^{-A\phi}\rb_{;M}\lb
e^{-A\phi}\rb_{;N}\td G^{MN}\nonumber\\
\phi\rightarrow\td\phi&=&\sqrt{2A}\phi
\label{eq:conftrans}
\eea
\end{minipage}

to obtain
\bea
S\rightarrow \td S&=&\frac{1}{2\kappa^2}\int
d^Dx\sqrt{\td{G}}\lc\td{R}-\td{G}^{MN}\nabla_M\td\phi\nabla_N\td\phi\rc\
,
\label{eq:eftta}
\eea
where $\td\phi$ is the
canonically-normalized dilaton, and we have ignored flux
contributions. We dimensionally reduce the action by integrating out
the extra dimensions~\cite{batwat,cghw}. To perform this last step we
consider a string-frame metric of the form~(\ref{eq:metric}), split
into $d$ large directions described by $g_{\mu\nu}$ and $n$ compact
directions described by $\gamma_{mn}$. For simplicity, we consider the geometry of the extra dimensions to be 
that of a torus, thus $R[\gamma_{mn}]=0$. We use the following
relations to isolate the scale-factor dependence on the
extra-dimensions~\cite{batwat,cghw,birdav}
\bea
\sqrt{-\td G}&=&\td b^n\sqrt{-\td g}\\
\td R=\td R[\td G_{MN}]&=&\td R[\td g_{\mu\nu}] -2\nc\td b^{-1}\td
g^{\mu\nu}\td\nabla_\mu\td\nabla_\mu\td b-\nc(\nc-1)\td
b^{-2}\td g^{\mu\nu}\td\nabla_\mu\td b\td\nabla_\nu \td b,
\eea
where, again, $R[\gamma_{mn}]=0$, $\nc$ and $\ti{b}(x^{\mu})$ are the number and scale factor corresponding to the extra
dimensions, and $\td{g}_{\mu\nu}$ is the  metric of the
non-compact directions. Since none of the terms in the action
depend explicitly on the coordinates from the $n$ extra dimensions, we
integrate over these directions to get the low energy effective action
of the $d+1$-dimensional theory
\bea
S_{eff}&=&\frac{V_{\nc}}{2\kappa^2}\int d^{d+1}x\sqrt{-\td g}\ls \lb\td b^{\nc} d\td R[\td g_{\mu\nu}] -2\nc\td b^{\nc-1}\td
g^{\mu\nu}\td\nabla_\mu\td\nabla_\mu\td b-\nc(\nc-1)\td
b^{\nc-2}\td g^{\mu\nu}\td\nabla_\mu\td b\td\nabla_\nu \td
b\rb \right.\nonumber\\
&&\hspace{3cm}\left.-\td b^{\nc}\td
  g^{\mu\nu}\td\nabla_\mu\td\phi\td\nabla_\nu\td\phi\rs,\label{eq:effbgc}
\eea
where $V_{\nc}\equiv \int d^{\nc}y\sqrt{\gamma}$ is the spatial volume of the
$n$ extra dimensions under unit scaling $(\td b =1)$.
\par
A second conformal transformation and field redefinition of the
action~(\ref{eq:effbgc}) is necessary to obtain the canonical
form of the Einstein-Hilbert action. The conformal transformation reuses the
identities~(\ref{eq:conftrans}) with 
\bea
\bg_{\mu\nu}=\td b^{\nc}\td g_{\mu\nu}\equiv e^{\sqrt{B}\td\psi}\td g_{\mu\nu}\ ,
\label{eq:efftrans}
\eea
resulting in
\bea
S_{eff}&=&\frac{V_n}{2\kappa^2}\int
 d^{d+1}x\sqrt{-\bg}\lb R[\bg_{\mu\nu}]-\bg^{\mu\nu}\bnabla_\mu
\td\psi\bnabla_\nu
\td\psi -\bg^{\mu\nu}\bnabla_\mu\tphi\bnabla_\nu\tphi\rb\
,\label{eq:effactnr}
\eea
where $B=\frac{d-1}{n(d+n-1)}$. Finally, the system is canonically
normalized by identifying the 4D~Planck mass as
$M_{p}^2\equiv\frac{V_n}{\kappa^2}$, and by rescaling the fields as
\bea
&&\psi=M_p\td\psi\  ,\ \varphi=M_p\,\td\phi\\
\Rightarrow S_{eff}&=&\int
d^{d+1}x\sqrt{-\bg}\lb\frac{M_p^2}{2}R[\bg_{\mu\nu}]-\frac{1}{2}\bg^{\mu\nu}\bnabla_\mu\psi\bnabla_\nu\psi
-\frac{1}{2}\bg^{\mu\nu}\bnabla_\mu\varphi\bnabla_\nu\varphi\rb
.\label{eq:effact}
\eea
\par 
The net effect of these transformations is to rescale the scale
factors and dilaton as
\bea
\sqrt{-G}&\rightarrow&\sqrt{-\bg}\,e^{D\sqrt{\frac{A}{2}}\frac{\varphi}{M_p}}\,e^{n\sqrt{B}\frac{\psi}{M_p}}\nonumber\\
a(t)&\rightarrow&
\bar{a}(t)=e^{\frac{\nc}{d-1}\sqrt{B}\frac{\psi}{M_p}}\,e^{-\sqrt{A/2}\frac{\varphi}{M_p}}a(t)\nonumber\\
b(t)&\rightarrow&\td{b}(t)=e^{\sqrt{B}\frac{\psi}{M_p}}\,e^{-\sqrt{A/2}\frac{\varphi}{M_p}}b(t)\nonumber\\
\phi(t)&\rightarrow&
\varphi(t)=\sqrt{2A}M_p\,\phi(t) ,
\eea
Employing the above expressions, we may now express the contribution
of a source behaving as
\bea
\rho&=&\rho_ia^{-d(1+\omega_i)}b^{-\nc(1+\homega_i)}
\eea
in the $D$ dimensional string frame,
 through the effective matter-action
\bea
S_{eff_m}&=&\int
d^{d+1}x\sqrt{-\bg}e^{-\alpha\sqrt{\frac{1}{2A}}\varphi}\bar\rho\nonumber\\
&=&\int d^{d+1}x\sqrt{-\bg} \rho_i\,e^{\lb-\homega_i+\frac{d}{d-1}\lb\omega_i-\frac1d\rb\rb
\sqrt{\frac{(d-1)\nc}{(d+\nc-1)}}\psi}\,e^{\lb-d\omega_i-\nc\homega_i+1-\alpha_i\frac{d+\nc-1}{2}\rb
\sqrt{\frac{1}{d+\nc-1}}\varphi}\,\bar{a}^{-d(1+\omega_i)} \nonumber\\
&\equiv&\int d^{d+1}x\sqrt{-\bg} \rho_i\,e^{2(\mu_i\varphi+\nu_i\psi)}
\eea
with $M_p=1$. The original
theory of dilaton gravity together with string and brane sources can
now be interpreted as a theory of Einstein gravity together with
sources, plus two scalar fields corresponding to the dilaton
($\varphi$) and the moduli field ($\psi$), this is the action of
equation~(\ref{eq:effact}). As well, the source term (equation~\ref{ap:mat}) now acts like an
effective potential for the two scalar fields. The inclusion of
different excited states will provide different effective potentials,
and this freedom can be exploited in the search for a moduli-stabilizing potential.

\section{Equations of state}
\label{app:exps}
In this section we derive the equations of state and the resultant
coefficients for the brane-gas effective potential. Using the
metric-ansatz~(\ref{eq:metric}), we derive the gas pressure from the
thermodynamic relation.
\bea
P_a&=&-\left.\frac{\delta E}{\delta V}\right|_{b=const.}
\eea
The volume is given by $V=\sqrt{-G_s}=a^db^\nc$, while energy
contributions are generically of the form $E=a^{j}b^{k}=\lb
a^d\rb^{\frac jd}\lb b^n\rb^{\frac kn}$, so that
\bea
\delta V&=&b^n\delta(a^d)+a^d\delta(b^n)\\
\delta
E&=&\frac{j}{d}\frac{a^{j}b^{k}}{a^d}\delta(a^d)+\frac{k}{n}\frac{a^{j}b^{k}}{b^n}\delta(b^n)\\
\Rightarrow P_a=-\left.\frac{\delta E}{\delta V}\right|_{b=const}&=&-\frac{j}{d}\frac{a^{j}b^{k}}{a^db^n}\frac{\delta(a^d)}{\delta(a^d)}=\omega\frac
EV=\omega \rho\\
\Rightarrow P_b=-\left.\frac{\delta E}{\delta V}\right|_{a=const}&=&-\frac{k}{n}\frac{a^{j}b^{k}}{a^db^n}\frac{\delta(b^n)}{\delta(b^n)}=\homega\frac
EV=\homega \rho\ ,
\eea
where we have made the identifications $\omega=-\frac{j}{d}$ and
$\homega=-\frac{k}{n}$. Thus $E=a^{j}b^{k}=a^{-d\omega}b^{-n\homega}$.
The existence of winding and momentum modes for strings is a
well-known result, and is the reason for the T--duality invariant spectrum of
closed strings. Finding an embedding with quantized
momentum modes of branes is less subtle because the T-dual of a
wrapped brane results in a wrapped brane, not a momentum mode. However, we
use the embedding described by Kaya~\cite{kaya_vol}, which results in
momentum modes in the compact direction with energy given by
\be
E_n=\frac{\lambda_n}{b(t)}\ ,
\ee  
where $\lambda_n$ is an unknown eigenvalue for the $n$'th momentum
mode (we choose $\lambda_n>0$). The corresponding pressure due to this
brane momentum-mode is
\be
P_n=\frac{\lambda_n}{b(t)}\ ,
\ee
which is a positive quantity.
\par

\section{D-string oscillator modes}
\label{app:dosc}
For completeness we consider the naive quantization of 
D-strings, in case these modes could affect the no-go result for
simultaneous stabilization of the dilaton and radion.
Ignoring the 2-form gauge field that couples to the D-string, the
spectrum of the D-strings looks identical to that of 
F-strings except for the replacement 
\be
M_s\longrightarrow M_s'=e^{-\varphi/2}M_s
\ee
The rescaling is again due to the dilaton coupling present in the DBI action for the D-strings. Provided we ignore $ N_{\rm{osc}}<0$ modes, again it is sufficient to consider only the ``pure'' modes. A straight forward computation yields the following source actions
\be
S_{\rm{D,mom}}=\int d^{D}x\ \sqrt{-\hat{g}}a^{-3}b^{-(\nc+1)}
\ee
\be
S_{\rm{D,osc}}=\int d^{D}x\ \sqrt{-\hat{g}}e^{-\varphi/2}a^{-3}b^{-\nc}
\ee
and
\be
S_{\rm{D,wind}}=\int d^{D}x\
\sqrt{-\hat{g}}e^{-\varphi}a^{-3}b^{-(\nc-1)}\ .
\ee

The D-string momentum modes looks identical to those of the F-string
momentum modes, and yield no new effect. The winding modes
are the same as those obtained for D1-branes,
which have already been considered. The only qualitatively new
contribution comes from the oscillatory D-string modes. 
Substituting $\al=1/2$ and
$\omo=0$ in (\ref{gen-munu}) one finds
\be
\te_{\rm{D,osc}}=-{\pi\over2}-{\pi\over 6}
\ee
Again, this fails to stabilize the runaway direction. 

\section{Solving the overshoot problem}
\label{app:zeta}
One can analytically estimate of what must be the initial ratio of energy
densities in the brane gas  and radion in order to solve the
overshoot problem. If the initial energy density of branes is much
smaller than the potential energy of the radion, the dynamical
equations will be given by \be
H^2\cong\3\left({\frac12\dot{\psi}^2}+Ee^{-2\al_1\psi}\right)
\ee 
and
\be
\ddot{\psi}+3H\dot{\psi}\cong 2\al_1 Ee^{-2\al_1\psi}
\ee
The radion rolls freely down the exponential potential and exact
solutions are known~\cite{liddle}:
\be
a\sim t^{1/2\al_1^2} \mx{ and } e^{\psi}\sim t^{1/\al_1}\sim a^{2\al_1}
\ee
Thus the energy densities of the brane and the radion redshift in this non-tracking phase as
\be
\rho_pe^{2\nu_p\psi}\sim a^{-(3-4\nu_p\al_1)}\mx{ while } V(\psi)\sim a^{-4\al_1^2}
\ee
Thus as long as 
\be
3-4\nu_p\al_1<4\al_1^2
\ee
the brane energy density will catch up with the potential energy of
the radion. We can calculate when this happens. 
The ratio of brane energy to radion potential energy is
\be
{\rho_pe^{2\nu_p\psi}\over V(\psi)}\sim a^{4\nu_p\al_1+4\al_1^2-3}
\ee
and we want that this ratio to be ${\cal O}(1)$, by the time the radion rolls to the minimum. Hence we need to start with an initial  ratio such that
\be
{\zeta}\equiv {\rho_{p0}e^{2\nu_p\psi_0}\over V(\psi_0)}=\left({a_0\over a_{min}}\right)^{4\nu_p\al_1+4\al_1^2-3}
\label{zeta-def}
\ee
where
\be
{V(\psi_0)\over V(\psi_{min})}=e^{-2\al_1\psi_0}=\left({a_0\over a_{min}}\right)^{-4\al_1^2}
\label{nontrack-scale}
\ee
From (\ref{zeta-def}) and (\ref{nontrack-scale}) we find
\be
\zeta=
\exp\left[2\al_1\psi_0\left(1-{3-4\nu_p\al_1\over 4\al_1^2}\right)\right]
\Ra 
\log_{10}\zeta \approx -0.43\left[2\al_1\psi_0\left(1-{3-4\nu_p\al_1\over 4\al_1^2}\right)\right]
\ee

\bibliographystyle{apsrmp}
\bibliography{rmp-sample}

\begin{thebibliography}{99}
\bibitem{branvafa}R.~H.~Brandenberger and C.~Vafa,
``Superstrings In The Early Universe,''
Nucl.\ Phys.\ B {\bf 316}, 391 (1989).
\bibitem{tseyvaf}A.~A.~Tseytlin and C.~Vafa, 
``Elements of string cosmology,''
{\it Nucl. Phys. B} {\bf
    372}, 443 (1992).
\bibitem{abe}S.~Alexander, R.~H.~Brandenberger and D.~Easson,
  ``Brane gases in the early universe,''
  Phys.\ Rev.\ D {\bf 62}, 103509 (2000)
  [arXiv:hep-th/0005212].
\bibitem{branwat_isot}S.~Watson and R.~H.~Brandenberger,
``Isotropization in brane gas cosmology,''
Phys.\ Rev.\ D {\bf 67}, 043510 (2003)
[arXiv:hep-th/0207168].
\bibitem{branwat_stab}
S.~Watson and R.~Brandenberger,
``Stabilization of extra dimensions at tree level,''
JCAP {\bf 0311}, 008 (2003)
[arXiv:hep-th/0307044].
\bibitem{egj}R.~Easther, B.~R.~Greene and M.~G.~Jackson,
``Cosmological string gas on orbifolds,''
Phys.\ Rev.\ D {\bf 66}, 023502 (2002)
[arXiv:hep-th/0204099].
\bibitem{GUT}
V.~S.~Kaplunovsky,
  ``Mass Scales Of The String Unification,''
  Phys.\ Rev.\ Lett.\  {\bf 55}, 1036 (1985);
  ``Couplings And Scales In Superstring Models,''
  Phys.\ Rev.\ Lett.\  {\bf 55}, 366 (1985);
  R.~Petronzio and G.~Veneziano,
  ``Constraints From String Unification,''
  Mod.\ Phys.\ Lett.\ A {\bf 2}, 707 (1987).
\bibitem{egjk1}
  R.~Easther, B.~R.~Greene, M.~G.~Jackson and D.~Kabat,
  ``String windings in the early universe,''
  JCAP {\bf 0502}, 009 (2005)
  [arXiv:hep-th/0409121].
\bibitem{anupam1}
R.~Danos, A.~R.~Frey and A.~Mazumdar,
  ``Interaction rates in string gas cosmology,''
  Phys.\ Rev.\ D {\bf 70}, 106010 (2004)
  [arXiv:hep-th/0409162].
\bibitem{5th}  C.~M.~Will,
  ``The confrontation between general relativity and experiment,''
  Living Rev.\ Rel.\  {\bf 4}, 4 (2001)
  [arXiv:gr-qc/0103036];
R.~Trotta, P.~P.~Avelino and P.~Viana,
  ``WMAP Constraints on varying $\alpha$ and the Promise of
Reionization,''
  Phys.\ Lett.\ B {\bf 585}, 29 (2004)
  [arXiv:astro-ph/0302295].

\bibitem{batwat}T.~Battefeld and S.~Watson,
``Effective field theory approach to string gas cosmology,''
JCAP {\bf 0406}, 001 (2004)
[arXiv:hep-th/0403075].
\bibitem{arakay}
S.~Arapoglu and A.~Kaya,
``D-brane gases and stabilization of extra dimensions in dilaton gravity,''
Phys.\ Lett.\ B {\bf 603}, 107 (2004)
[arXiv:hep-th/0409094].
\bibitem{brustein}
  R.~Brustein and P.~J.~Steinhardt,
  ``Challenges for superstring cosmology,''
  Phys.\ Lett.\ B {\bf 302}, 196 (1993)
  [arXiv:hep-th/9212049].
\bibitem{kklt}
 S.~Kachru, R.~Kallosh, A.~Linde and S.~P.~Trivedi,
  ``De Sitter vacua in string theory,''
  Phys.\ Rev.\ D {\bf 68}, 046005 (2003)
  [arXiv:hep-th/0301240].
\bibitem{bek}R.~Brandenberger, D.~A.~Easson and D.~Kimberly,
``Loitering phase in brane gas cosmology,''
Nucl.\ Phys.\ B {\bf 623}, 421 (2002)
[arXiv:hep-th/0109165].
\bibitem{kaya_vol}
A.~Kaya,
``Volume stabilization and acceleration in brane gas cosmology,''
JCAP {\bf 0408}, 014 (2004)
[arXiv:hep-th/0405099];
T.~Rador,
  arXiv:hep-th/0504047; 

  arXiv:hep-th/0502039.
\bibitem{zhuk}
U.~Gunther, S.~Kriskiv and A.~Zhuk,
  Grav.\ Cosmol.\  {\bf 4}, 1 (1998)
  [arXiv:gr-qc/9801013]; 
U.~Gunther and A.~Zhuk,
  Class.\ Quant.\ Grav.\  {\bf 15}, 2025 (1998)
  [arXiv:gr-qc/9804018].

\bibitem{bercli}
A.~J.~Berndsen and J.~M.~Cline,
  ``Dilaton stabilization in brane gas cosmology,''
  Int.\ J.\ Mod.\ Phys.\ A {\bf 19}, 5311 (2004)
  [arXiv:hep-th/0408185].
\bibitem{gubpee}
S.~S.~Gubser and P.~J.~E.~Peebles,
  ``Structure formation in a string-inspired modification of the cold dark
  matter model,''
  Phys.\ Rev.\ D {\bf 70}, 123510 (2004)
  [arXiv:hep-th/0402225].
\bibitem{Watson}
S.~Watson,
``Moduli stabilization with the string Higgs effect,''
Phys.\ Rev.\ D {\bf 70}, 066005 (2004)
[arXiv:hep-th/0404177].
\bibitem{subodh}
S.~P.~Patil and R.~H.~Brandenberger,
``The cosmology of massless string modes,''
  arXiv:hep-th/0502069;
S.~P.~Patil,
  ``Moduli (dilaton, volume and shape) stabilization via massless F and D
  string modes,''
  arXiv:hep-th/0504145.

\bibitem{prashanth}
T.~Biswas and P.~Jaikumar,
  JHEP {\bf 0408}, 053 (2004)
  [arXiv:hep-th/0407063]; 
Int.\ J.\ Mod.\ Phys.\ A {\bf 19}, 5443 (2004).
\bibitem{GKP}
S.~B.~Giddings, S.~Kachru and J.~Polchinski,
  ``Hierarchies from fluxes in string compactifications,''
  Phys.\ Rev.\ D {\bf 66}, 106006 (2002)
  [arXiv:hep-th/0105097].
\bibitem{deAlwis}
  S.~P.~de Alwis,
  ``Effective potentials for light moduli,''
  arXiv:hep-th/0506266.
\bibitem{quevedo}
  J.~P.~Conlon, F.~Quevedo and K.~Suruliz,
  ``Large-volume flux compactifications: Moduli spectrum and D3/D7 soft
  supersymmetry breaking,''
  arXiv:hep-th/0505076.
\bibitem{KS}
  I.~R.~Klebanov and M.~J.~Strassler,
  ``Supergravity and a confining gauge theory: Duality cascades and
  $\chi$SB-resolution of naked singularities,''
  JHEP {\bf 0008}, 052 (2000)
  [arXiv:hep-th/0007191].
\bibitem{anupam2}
  T.~Biswas and A.~Mazumdar,
  ``Can we have a stringy origin behind $\Omega_\Lambda(t)$ 
proportional to   $\Omega_m(t)$?,''
  arXiv:hep-th/0408026;
T.~Biswas, R.~Brandenberger, A.~Mazumdar and T.~Multamaki,
  arXiv:hep-th/0507199.
\bibitem{biswas}
  T.~Biswas, R.~Brandenberger, D.~A.~Easson and A.~Mazumdar,
  ``Coupled inflation and brane gases,''
  Phys.\ Rev.\ D {\bf 71}, 083514 (2005)
  [arXiv:hep-th/0501194].
\bibitem{dealwis}
  R.~Brustein, S.~P.~de Alwis and P.~Martens,
  ``Cosmological stabilization of moduli with steep potentials,''
  Phys.\ Rev.\ D {\bf 70}, 126012 (2004)
  [arXiv:hep-th/0408160].
\bibitem{ovrut}
G.~Huey, P.~J.~Steinhardt, B.~A.~Ovrut and D.~Waldram,
  ``A cosmological mechanism for stabilizing moduli,''
  Phys.\ Lett.\ B {\bf 476}, 379 (2000)
  [arXiv:hep-th/0001112].
\bibitem{sorbo}
N.~Kaloper, J.~Rahmfeld and L.~Sorbo,
``Moduli entrapment with primordial black holes,''
Phys.\ Lett.\ B {\bf 606}, 234 (2005)
[arXiv:hep-th/0409226].
\bibitem{Easson:2005ug}
  D.~A.~Easson and M.~Trodden,
  ``Moduli stabilization and inflation using wrapped branes,''
  arXiv:hep-th/0505098.
\bibitem{cghw}S.~M.~Carroll, J.~Geddes, M.~B.~Hoffman and R.~M.~Wald,
``Classical stabilization of homogeneous extra dimensions,''
Phys.\ Rev.\ D {\bf 66}, 024036 (2002)
[arXiv:hep-th/0110149].
\bibitem{birdav}Birrell,~N.~D., Davies,~P.~C., ``Quantum Fields in
    Curved Space'', Cambridge University Press, 1982.  
\bibitem{liddle} 
E.~J.~Copeland, A.~R.~Liddle and D.~Wands,
  Phys.\ Rev.\ D {\bf 57}, 4686 (1998)
  [arXiv:gr-qc/9711068].
\end{thebibliography}

\end{document}